\documentstyle[11pt]{article} 
\def\build#1_#2^#3{\mathrel{
\mathop{\kern0pt#1}\limits_{#2}^{#3}}}

\def\ga{\mathrel{\mathpalette\fun >}}
\def\fun#1#2{\lower3.6pt\vbox{\baselineskip0pt\lineskip.9pt
        \ialign{$\mathsurround=0pt#1\hfill##\hfil$\crcr#2\crcr\sim\crcr}}}

\def\repx{|e\left ({\bf x},\,{\bf p}\right )\rangle}
\def\lepx{\langle e\left ({\bf x},\,{\bf p}\right ) |}
\def\rpx{|e\left ( x,\, p\right )\rangle}
\def\lpx{\langle e\left ( x,\, p\right ) |}
\def\bx{{\bf x}}
\def\bp{{\bf p}}
\def\ben{\begin{equation}}
\def\be#1{\begin{equation}\label{eq:#1}}
\def\ee{\end{equation}}
\def\EC#1{(\ref{eq:#1})}

\def\ldb{\lambda_{\rm de\,B}}

\begin{document}

\begin{center}   
\vspace{.2in}

{\Large \bf Modeling Collisionless Matter in General Relativity:\\
\bigskip
A New Numerical Technique}\\
\vspace{.2in}   
  
Lawrence M. Widrow \\
\vspace{.1in}
\medskip
$^1${\it Department of Physics\\
Queen's University, Kingston, Canada, K7L 3N6}\
\end{center}
  
\vspace{.3in}   
  

\abstract{

We propose a new numerical technique for following the evolution of a
self-gravitating collisionless system in general relativity.  Matter
is modeled as a scalar field obeying the coupled Klein-Gordon and
Einstein equations.  A phase space distribution function, constructed
using covariant coherent states, obeys the relativistic Vlasov
equation provided the de Broglie wavelength for the field is very much
smaller than the scales of interest.  We illustrate the method by
solving for the evolution of a system of particles in a static,
plane-symmetric, background spacetime.
   
}
\noindent 

\newpage   
\pagestyle{plain}   

\section{Introduction}

There now exists techniques to calculate numerically the dynamical
evolution of matter in general relativity.  For the most part
research has focused on three very different types of systems; fluids,
scalar fields, and collisionless matter.  In this work we develop and
exploit a connection between the latter two.

A collisionless system with a large number of particles is treated
generally as a continuous fluid in phase space.  The fluid is described
by a distribution function $f$ which gives, for each region of the
system, the density of particles and their distribution in velocity
space.  $f$ obeys the Vlasov equation with an appropriate force law
which, in general relativity, is given by the geodesic and Einstein
equations.  Shapiro and Teukolsky [1] have developed a computational
method for handling collisionless relativistic systems that combines
N-body techniques and numerical relativity.  An alternative approach
[2] is to follow the evolution of $f$ directly in phase space by
solving the coupled Einstein and Vlasov equations.  Applications
include violent relaxation and the collapse to a black hole of an
unstable relativistic star cluster.

In this work we show that a massive scalar field obeying the coupled
Klein-Gordon and Einstein equations provides an alternative model for
collisionless relativistic systems that can be readily adapted to the
computer.  The technique makes use of what is essentially the 
scalar field analogue of geometric optics in general relativity [3].
The spirit and methodology is similar to that
of Widrow and Kaiser [4] who showed that a field obeying the coupled
Schr\"odinger and Poisson equations could be used to model
collisionless, nonrelativistic matter.

Scalar fields in general relativity have been studied for some time,
albeit for entirely different reasons.  For the past fifteen years the
motivation for much of this work has been to understand the
inflationary universe paradigm.  For example, numerical simulations of
inhomogeneous scalar field cosmologies have been studied in an attempt
to understand the onset of inflation [5].  Recently, there has been
great deal of interest in the gravitational collapse to a black hole
of a self-gravitating massless scalar field.  This is due primarily to
the discovery by Choptuik [6] that such systems exhibit scaling
behaviour and critical phenomena.

Again our interest here is in simulations of collisionless matter.
For our purposes previous investigations of scalar fields are
important in that they exhibit the numerical techniques used to follow
their dynamics.  Our discussion does suggest that the
results found by Choptuik might also apply to collisionless matter.
Unfortunately the simulations that have shown scaling behaviour and
critical phenomena involve massless scalar fields whereas
our work requires a nonzero mass.

We model a collisionless relativistic 
system by a field configuration $\phi(x)$ where $\phi$ is a complex
scalar field obeying the Klein-Gordon equation in curved space:

\be{kggr} 
\left (g^{\mu\nu}\nabla_\mu\partial_\nu~+~ \frac{m^2}{\hbar^2}\right
)\phi(x)~=~0 ~.
\ee 

\noindent (The metric $g_{\mu\nu}$ has signature
$(+\,-\,-\, -)$ and we set $c=1$.)  $m/\hbar$ is a model parameter
which must be large enough to guarantee that geometric optics (or
rather the scalar field analogue of geometric optics) applies.  In
particular we require that throughout the system $\ldb\ll {\cal L}$
and $\ldb\ll {\cal R}$ where ${\cal L}$ is the scale over which the
density and velocity field of the system (i.e., distribution function)
vary, ${\cal R}$ is the typical radius of curvature for the spacetime,
and $\ldb\sim |\phi/\nabla\phi |$ is the typical de Broglie
wavelength for the field.  Locally $\phi$ can be regarded as the
superposition of plane waves propagating along geodesics [3].  The
amplitude of the field in a given region of the system tells us
something about the density of particles while information about the
velocity distribution of the particles is encoded in the phase structure
of the field.

The discussion above suggests the following prescription for
constructing a distribution function ${\cal F}={\cal F}(x,p)$ from the
field $\phi$: Fourier transform the field in a region of size $\eta$
($\ldb\ll\eta$ and $\eta\ll {\cal L},{\cal R}$) centered on a
particular point in phase space and then take the absolute square.
This is what is done in [4] for a nonrelativistic Schr\"odinger field
$\psi=\psi({\bf x},t)$.  There ${\cal F}$ is constructed using
coherent states: ${\cal F}(\bx,\,\bp)~=~|\lepx\psi\rangle |^2$ where
$\repx$ is a localized state centered on $\bx$ with average momentum
$\bp$.  $\repx$ is taken to be a Gaussian (minimum uncertainty)
wavepacket,

\be{nrwp}
\lepx {\bf
x}'\rangle=\left (\frac{1}{2\pi\hbar}\right )^{3/2}
\left (\frac{1}{\pi\eta^2}\right )^{3/4}
e^{{-\left (\bx-\bx'\right
)^2/2\eta^2-i\bx'\cdot\bp/\hbar }}\, ,
\ee

\noindent having width $\eta$ in position space and $\hbar/\eta$ in momentum
space.  The normalization is chosen so that $\int d^3 x\int d^3 p
{\cal F}(x,p)=1$ provided $\int d^3x |\psi(x)|^2 =1$.  It is
straightforward to show that $\cal F$ obeys approximately the
nonrelativistic Vlasov equation provided $\ldb\ll\eta\ll {\cal L}$.

The coherent-state representation for a scalar field in special
relativity is again based on a set of state vectors $\rpx$ localized
in both position and momentum.  We choose minimum uncertainty
wavepackets [7]:

\be{grwp}
\lpx p'\rangle~=~
\left (\frac{\eta^2}{\pi\hbar^2}\right )^{3/4}
e^{-ix\cdot p'/\hbar-p\cdot p'\eta^2/\hbar^2}
\ee

\noindent where
$p=(p_0,\bp )$ with $p_0>|\bf p|$ and $a\cdot b\equiv a_0 b_0 -{\bf
a}\cdot {\bf b}$.  These wavepackets are centered about ${\bf x}$ 
and move with average momentum ${\bf p}$.

The extension to general relativity is most easily
accomplished using Riemann normal coordinates (RNCs) and is similar to
discussions found in the literature [8] of the Wigner function in
curved space.  In a RNC system curved space closely resembles flat
space in the neighborhood of a particular point taken to be the
origin.  The metric can then be written as a Taylor series about the
flat space metric $\eta_{\mu\nu}$ in increasing orders of the Riemann
tensor $R_{\mu\nu\rho\sigma}$:
$g_{\mu\nu}~=~\eta_{\mu\nu}~+~\frac{1}{3}R_{\mu\rho\nu\sigma} y^\rho
y^\sigma~+\dots~$.  In RNCs the phase factor $x\cdot p = \eta_{\mu\nu}
x^\mu p^\nu + `{\rm post-geometric~optics~corrections}'$ [3].  $\phi$
in the curved space coherent-state representation is then

\be{grcs}
\lpx \phi\rangle~=~\int dV_p\,
e^{-ix\cdot p'/\hbar-p\cdot p'\eta^2/\hbar^2}\,\hat\phi(p')
\ee

\noindent where 

\be{defvp}
dV_p~\equiv~ \frac{dp_0dp_1dp_2dp_3}{\left (2\pi\right )^4}
2\pi\delta\left (p^2-m^2\right )\Theta\left (p_0\right )
\ee

\noindent is the momentum space volume
element on the mass shell, $\Theta(x)=1$ for $x\ge 0$ and zero
otherwise, and $\hat\phi(p)$ is the Fourier transform of $\phi(x)$:

\be{ftofphi}
\phi(x)~=~\int dV_p\, e^{-ip\cdot x/\hbar}\,\hat\phi(p)~.
\ee

\noindent
Note that $\lpx\phi\rangle$ reduces to $\lepx\psi\rangle$ for $|{\bf
p}|\ll m$.  In practice, one integrates over $p_0$ making use of the
identity $\int dV_p = d^3p/\left ( (2\pi)^3 2p^0\right )$.
For example
\begin{eqnarray}
\label{grf}
{\cal F}(x,p)~&=&\int dV_p' dV_p''e^{ix\cdot\left (p'-p''\right )/\hbar
-p\cdot\left (p'+p''\right )\eta^2/\hbar^2}
\hat\phi(p')\hat\phi^*(p'')\\
&=&~\int \frac{d^3p'}{\left (2\pi\right )^3 2p'^0}
\frac{d^3p''}{\left ( 2\pi\right )^3 2p''^0}
e^{ix\cdot\left (p'-p''\right )/\hbar
-p\cdot\left (p'+p''\right )\eta^2/\hbar^2}
\hat\phi(p')\hat\phi^*(p'').
\end{eqnarray}
In the second of these expression the integrand is evaluated
``on mass-shell'', i.e., with $p'^0=\sqrt{{\bf p'}\cdot{\bf p'}
+m^2}$ and $p''0=\sqrt{{\bf p''}\cdot{\bf p''}
+m^2}$.

Consider the function $\sigma(p) \equiv\delta\left ( p^2
-m^2\right)\Theta \left (p_0\right )\hat\phi(p)$.  From
Eq.\,\EC{kggr} and the definition of the RNCs we have

\be{eofmforsigma}
\left (
\frac{m^2-p^2}{\hbar^2}~-~
\frac{1}{3}R^{\mu~\nu}_{~\rho~\sigma}
\frac{\partial^2}{\partial p_\rho\partial p_\sigma}p_\mu p_\nu
~-~\frac{2}{3}R^\mu_{~\rho}\frac{\partial}{\partial p_\rho}p_\mu
\right )\sigma(p)~\equiv~{\cal O}(p)\sigma(p)~=~0  
\ee

\noindent 
from which follows immediately the identity

\be{startproof}
0~=~\int d^4p' d^4p'' e^{ix\cdot\left (p'-p''\right )/\hbar
-p\cdot\left (p'+p''\right )\eta^2/\hbar^2}
\left ({\cal O}(p')\sigma(p')\right )\sigma^*(p'')~.
\ee

\noindent A straightforward but tedious calculation leads to a complex equation
for $\cal F$ which can be split into real and imaginary parts:

\be{realeofm} 
\left (\frac{m^2-p^2}{\hbar^2}~+~\dots\right ){\cal
F}~=~0 
\ee 

\be{imageofm} 
\left
(\eta^{\mu\nu}p_\nu\frac{\partial}{\partial x^\mu}
~+~\frac{1}{3}R^{\mu~\nu}_{~\rho~\sigma}x^\sigma p_\mu p_\nu
\frac{\partial}{\partial p_\rho}~+~\dots\right ){\cal F}~=~0 ~.
\ee 

\noindent The ``$\dots$'' in these 
equations refer to terms higher order in the $\eta/L$, $\lambda_{\rm
deB}/\eta$ and $\eta/{\cal R}$ (i.e. post-geometric optics corrections).  
The first of these equations tells us that the
distribution function is peaked on the $p_\mu p^\mu=m^2$ (mass shell) surface.  The
second equation is, to leading order, the relativistic Vlasov equation
in a RNC system.

For self-gravitating collisionless matter, $g_{\mu\nu}$ depends on
$f$ through the Einstein equations,
$R_{\mu\nu}-\frac{1}{2}g_{\mu\nu}R=8\pi GT_{\mu\nu}$, where the
stress-energy tensor $T_{\mu\nu}$ is given by

\be{deft}
T_{\mu\nu}~=~\int\frac{d^4 p}{\sqrt{-g}}\, p_\mu p_\nu\,f(x,p)~.
\ee

\noindent We can calculate $T_{\mu\nu}$ for our model by substituting $\cal F$
for $f$ in Eq.\,\EC{deft}.  This is, however, computationally expensive
(one must update $g$ at each timestep) and not really necessary.
Consider instead the expression

\begin{eqnarray}
\label{easyt}
T_{\mu\nu}~&=&~\partial_\mu\phi\partial_\nu\phi^* \\
&=&~\int dV_{p'} dV_{p''} e^{i(p''-p')\cdot x/\hbar}p_\mu' p_\nu'
\hat\phi(p')\hat\phi^*(p'')~.
\end{eqnarray}

\noindent Eq.\,\EC{deft} is what one 
would obtain if this expression were smoothed over a length scale
$\sim\eta$ and since the quantity of interest, $g\sim\nabla^{-2} T$,
is itself smoother than $T$ we can use either expression.
Eq.(12) is just the usual expression for the stress energy
tensor of a scalar field, save for a term proportional to
$\partial_\mu\phi\partial^\mu\phi^*-m^2\phi\phi^*$.  But this term is
negligible in the geometric optics limits [9].  

Evidently a self-gravitating scalar field behaves like collisionless
matter and can be represented in phase space using the coherent-state
representation described above provided $ldb\ll\eta\ll {\cal
L},\,{\cal R}$.  Of course to follow the field properly we must choose
a grid spacing that is somewhat less than $\ldb$.  In addition the
timestep must be short enough to follow the rapid evolution of the
phase factor for $\phi$.  The accuracy of a simulation is limited by
constraints on memory and CPU time.  The situation is similar to what
is encountered in particle methods where a finite number of particles
is used to provide a statistical description of the distribution
function.  One can improve the accuracy of a given simulation by
increasing the number of particles but again this comes at the cost of
memory and CPU time.

As an illustration of our method we consider ``particles'' in a static
and plane symmetric background spacetime.  For simplicity, we choose
the line element $ds^2= A^2dt^2-A^{-2}dz^2$ with $A^2=\exp{\left
(\omega^2 z^2\right )}$ so that nonrelativistic particles near the
origin will execute simple harmonic motion.  It is convenient to write
our equations in terms of the dimensionless quantities $\zeta=\omega
z$, $\tau=\omega t$, $\kappa=p/m$ and ${\cal L}=m/\hbar\omega$.
Eq.\,\EC{kggr}, for example, becomes

\be{scaledkg} 
\frac{\partial^2\phi}{\partial t^2}~-~
e^{2z^2}\frac{\partial^2\phi}{\partial z^2}~=~ e^{z^2}{\cal L}^2\phi~.
\ee 

\noindent $\lambda_{\rm de B}$ should be made as small as possible
(this is done by increasing ${\cal L}$) but no smaller than several
grid spacings.  With this in mind, we choose ${\cal L}\simeq 0.1N$ and
$\eta\simeq {\cal L}^{1/2}$ where $N$ is the number of gridpoints used
in the simulation.  There are roughly $N^{1/2}$ resolution elements in
both position and momentum space which is roughly what one would
expect for an N-body simulation with $N$ particles.

Consider first an initially cold (zero velocity dispersion)
distribution of particles.  The field configuration at $\tau=0$ is
taken to be $\phi(\zeta)= \exp{\left (-\zeta^2/\zeta_0^2\right )}$ and
$\partial\phi/\partial t=-i{\cal L}\phi$ with $\zeta_0=0.5$.  The
corresponding ${\cal F}$ is shown in the top left panel of Fig 1.  The
phase space configuration at $\tau=4\pi$ (i.e., time when particles
near the center of the distribution have made two complete orbits) is
shown in the bottom left panel.  For comparison, the results of an
N-body calculation are shown in the right-hand panels.  As expected,
particles near the center of the distribution execute simple harmonic
motion while particles initially at $z\ga 1$ follow anharmonic,
relativistic orbits in phase space.

We next consider a system of ``hot'' particles.  In particular, we
assume the system is initially described by a truncated isothermal
distribution function:
$f(\epsilon)=\Theta(\epsilon)\exp{(\epsilon/T')}$ where
$\epsilon\equiv p_0/E_{\rm max}$ and $T'\equiv T/E_{\rm max}$.  
For our simulation, we take $E_{\rm max}=2m$ and $T=m$.  $p_0$
is conserved for each particle in the system and therefore
$f(\epsilon)$ is constant.
 
The initial wavefunction for the Klein-Gordon simulation is taken to be

\be{initwf}
\phi(x)~=~\frac{1}{N}\sum_{i=1}^N
\langle e\left (x_i,\,p_i\right )| x\rangle
\ee

\noindent where the pairs $(x_i,p_i)$ are chosen at random from the
distribution function.  This is reminiscent of N-body simulations
where a statistical representation of the initial distribution
function is constructed by specifying the positions and velocities of
$N$ super-particles.  Much of the art in N-body simulations is in
calculating the forces and various methods (e.g., particle-particle
with smoothing, particle-mesh) are used to deal with such problems as
artificial two-body relaxation.  Here, the discrete particles are
replaced by wavepackets and so smoothing is built in.  The pros and
cons of our method, as compared with N-body techniques, will be
explored in a future publication [10].

Fig.\,2 shows the distribution function ${\cal F}$ for $\tau=0$ and
$\tau=8\pi$; Fig.\,3 gives ${\cal F}(p_0)$ (calculated from ${\cal
F}(x,p)$ using a simple binning procedure) again for $\tau=0$ and
$\tau=8\pi$.  While individual features of ${\cal F}$ change (this
also occurs in particle realizations of $f$) the overall structure
of the distribution function remains constant.

This work illustrates that simulations of collisionless matter in general
relativity can be done by following the dynamics of a massive scalar field
and in so doing, bridges the gap between two very separate branches of
numerical relativity.  Applications to self-gravitating systems with
more complicated geometries are straightforward only because of the large
volume of work on relativistic scalar fields that already exists.

\bigskip

{\bf \noindent Acknowledgements:} It is a pleasure to thank George
Davies and Kayll Lake for useful conversation.  This work was supported by a grant
from the Natural Science and Engineering Research Council of Canada.

\vfill\eject

\vskip 2cm
{\Large \noindent \bf References}
\vskip 0.5cm

\noindent[1] S. L. Shapiro and S. A. Teukolsky, Astrophys. J. {\bf 307}, 575 (1986)
and references therein.

\noindent[2] F. A. Rasio, S. L. Shapiro, and S. A. Teukolsky, Astrophys. J. {\bf 344},
146 (1989).  

\noindent[3] See, for example C. W. Misner, K. S. Thorne, and J. A. Wheeler,
Gravitation, W. H. Freeman \& Co., San Francisco, 1970, Section 22.5

\noindent[4] L. M. Widrow and N. Kaiser,
Astrophys. J. Lett. {\bf 416} L71 (1993).  

\noindent[5] See, e.g.,
D. S. Goldwirth and T. Piran, Phys. Rev. D {\bf 40}, 3263 (1989).

\noindent[6] M. W. Choptuik, Phys. Rev. Lett. {\bf 70}, 9 (1993).

\noindent[7] G. Kaiser, J. Math. Phys. {\bf 18}, 952 (1977).

\noindent[8] E. Calzetta, S. Habib, and  B. L. Hu, Phys. Rev. D {\bf 37},
2901 (1988) and references therein.

\noindent[9] S. R. de Groot, V. A. van Leeuwen, and Ch. G. van Weert,
{\it Relativistic Kinetic Theory} (North Holland Publishing Co., Amsterdam)
(1980) pp 70-73.
 
\noindent[10] G. Davies and L. M. Widrow, {\it in preparation}.

\bigskip

\vskip 2cm
{\Large \noindent \bf Figure Captions}
\vskip 0.5cm

\noindent FIG.1.  Evolution in phase space of an initially cold
distribution of particles.  The system is set in a fixed background
spacetime as described in the text.  4096 grid points are used for the
Klein-Gordon simulation.  Left-hand panels give the Klein-Gordon
distribution function ${\cal F}$ while right-hand panels give the
corresponding N-body distribution function $f$.

\bigskip

\noindent FIG.2.  Evolution of a ``hot'' system in phase space.
The initial distribution function is that of a truncated isothermal
sphere and is shown in the left panel.  The distribution function for
$\tau=8\pi$ is shown in the right panel.  2048 grip points are used
for the simulation.

\bigskip

\noindent FIG.3.  Distribution function ${\cal F}(E)$.  ${\cal F}(E)$
is calculated from the ${\cal F}(x,p)$ of Fig.2 using a simple binning
procedure.  Solid curve is the initial ${\cal F}(E)$; dashed curve is
the ${\cal F}(E)$ at $\tau=8\pi$; dotted curve is the  ${\cal F}(E)$
used to construct the initial wavefunction.

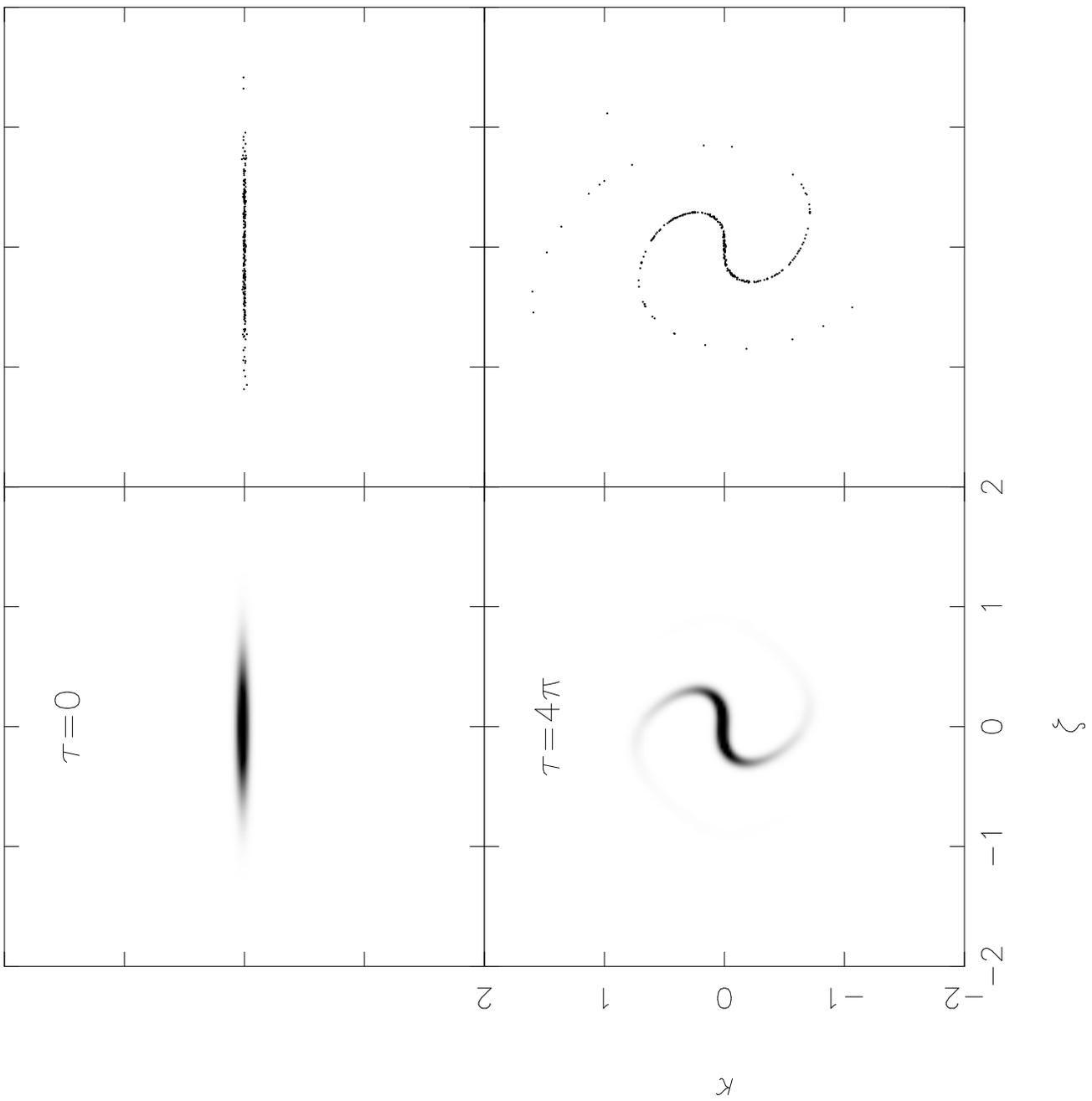
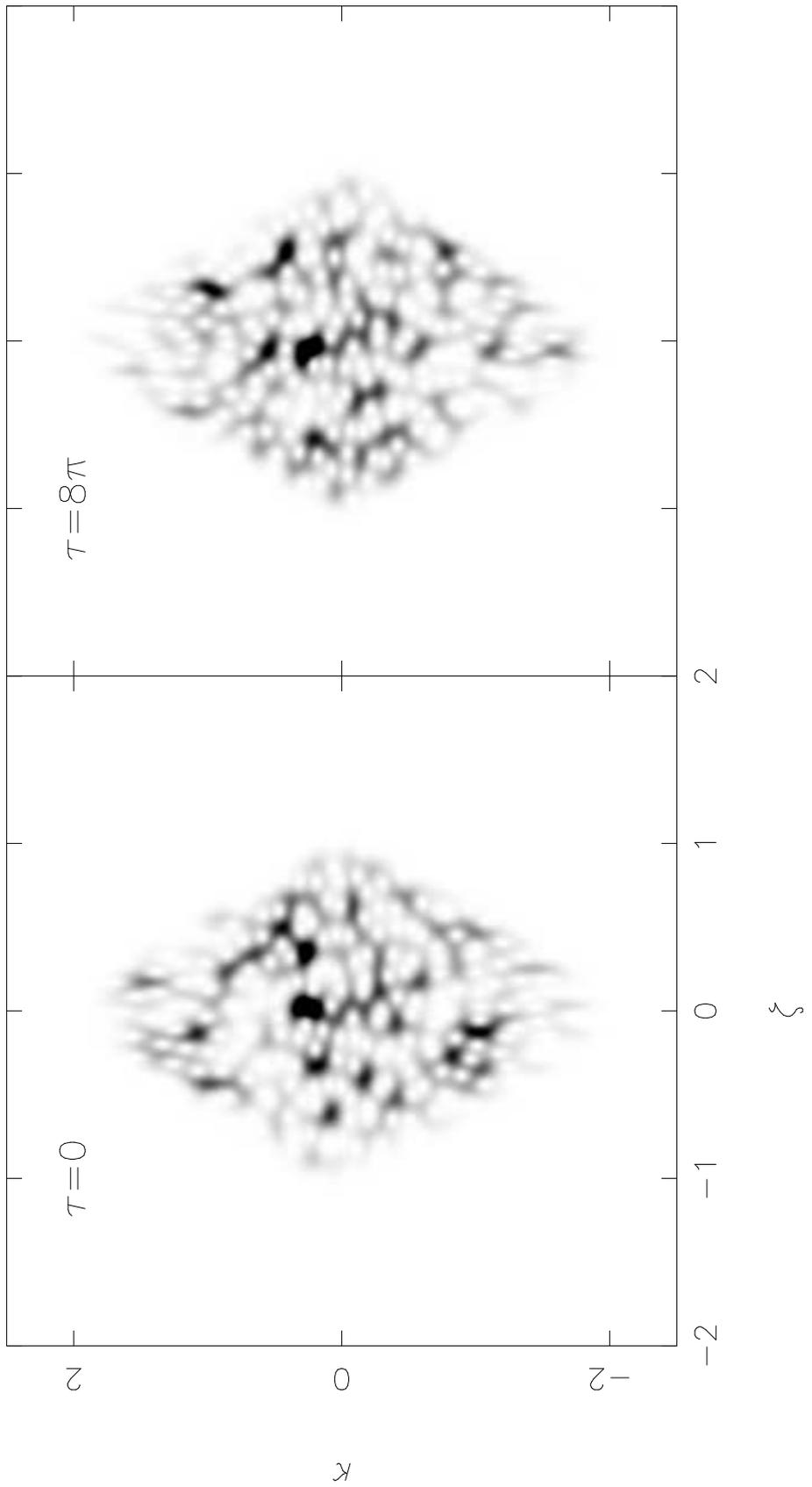
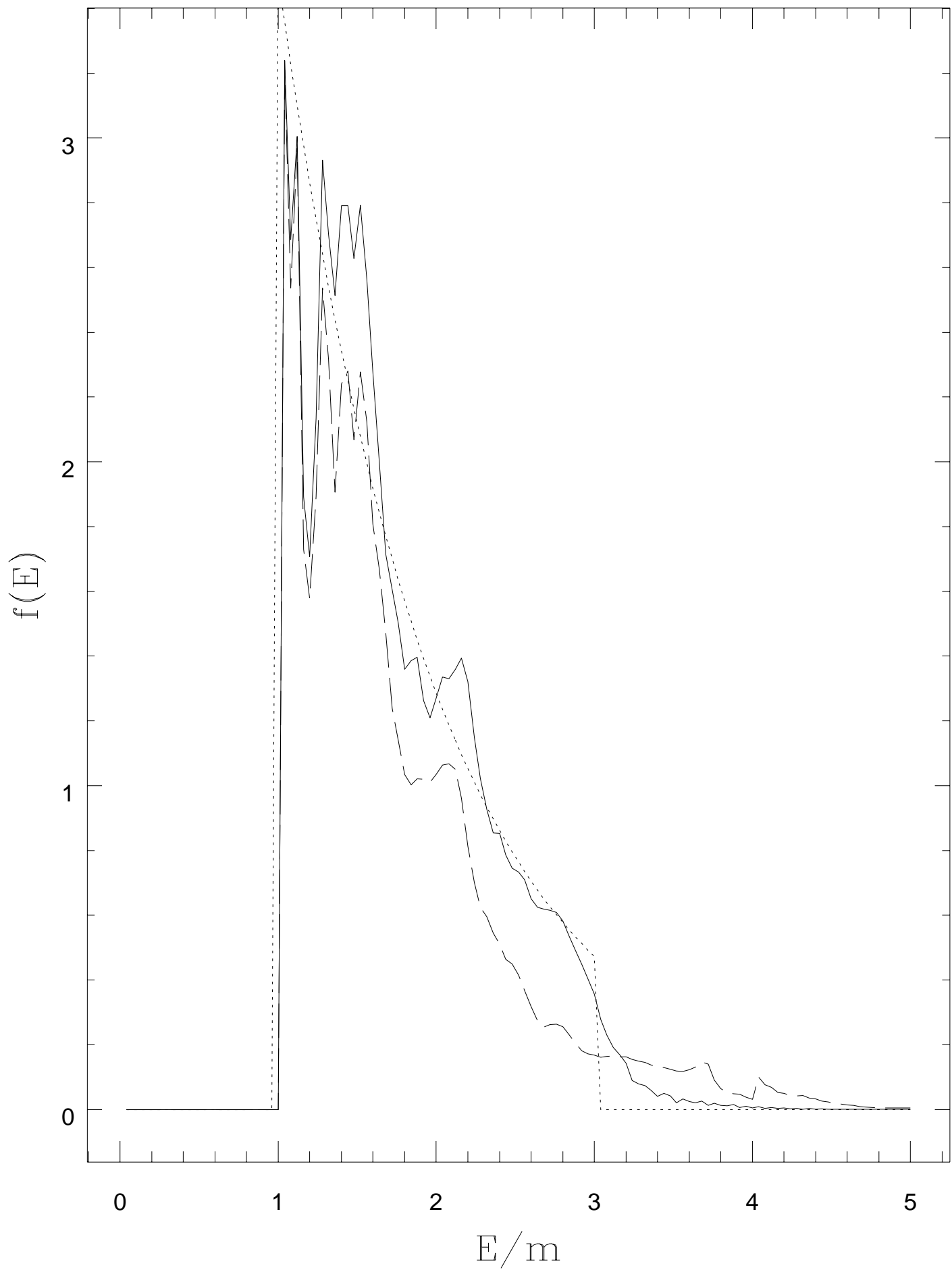
\end{document}